\pdfoutput=1

\documentclass[11pt]{article}

\usepackage{acl}

\usepackage{times}
\usepackage{latexsym}

\usepackage[T1]{fontenc}

\usepackage[utf8]{inputenc}

\usepackage{microtype}
\usepackage[nobiblatex]{xurl}
\usepackage{booktabs} 
\usepackage{multirow}
\usepackage{graphicx}
\usepackage{subcaption}
\usepackage{makecell}
\usepackage{xcolor}
\definecolor{repubred}{RGB}{232, 27, 35}
\definecolor{demblue}{RGB}{10, 83, 228}

\title{\textsc{CommunityLM}: Probing Partisan Worldviews from Language Models}

\author{Hang Jiang, Doug Beeferman, Brandon Roy, Deb Roy \\
  MIT Center for Constructive Communication \\
  MIT Media Lab\\
  \texttt{\{hjian42,dougb5,bcroy,dkroy\}@media.mit.edu} \\
  }

\begin{document}
\maketitle
\begin{abstract}
As political attitudes have diverged ideologically in the United
States, political speech has diverged linguistically.  The
ever-widening polarization between the US political parties is
accelerated by an erosion of mutual understanding between them.  We
aim to make these communities more comprehensible to
each other with a framework that probes community-specific responses
to the same survey questions using community language models
(\textsc{CommunityLM}). In our framework we identify committed
partisan members for each community on Twitter and fine-tune LMs on
the tweets authored by them.  We then assess the worldviews of the two
groups using prompt-based probing of their corresponding LMs, with
prompts that elicit opinions about public figures and groups surveyed
by the American National Election Studies (ANES) 2020 Exploratory
Testing Survey.  We compare the responses generated by the LMs to the
ANES survey results, and find a level of alignment that greatly
exceeds several baseline methods.  Our work aims to show that we can
use community LMs to query the worldview of any group of people given a
sufficiently large sample of their social media discussions or media
diet.

\end{abstract}

\section{Introduction}

Political polarization is a prominent component of politics in the United States \cite{poole1984polarization,mccarty2016polarized,heltzel2020polarization}. Previous studies have shown growing polarization in social media \cite{bail2018exposure,demszky2019analyzing,darwish2019quantifying} and substantial partisan and ideological differences in media diet \cite{bozell2004weapons,gil2012selective,hyun2016agenda}. \citet{li2017speaking} show that partisanship makes reliable predictions about an individual's word understanding. \citet{notthesame} used modern machine-translation techniques to demonstrate that the left and right communities use English words differently. \citet{milbauer-etal-2021-aligning} extended the method to uncover worldview and ideological differences between 32 Reddit communities. 
These studies are word-level analyses based on Word2vec word embeddings, and none of them use pre-trained language models. 


Prompting is a standard technique to make pre-trained language models generate texts conditioned on prompts. Recent work has shown that, through prompt engineering, pre-trained language models can achieve good zero-shot performance on NLP tasks from sentiment classification to reading comprehension \cite{radford2019language,brown2020language} and mine factual or commonsense knowledge \cite{petroni-etal-2019-language,davison-etal-2019-commonsense,jiang-etal-2021-know,talmor-etal-2020-olmpics}. Through prompting, \citet{palakodety2020mining} used a fine-tuned BERT \cite{devlin-etal-2019-bert} model with fill-in-the-blank cloze statements to mine insights and compare prediction differences between Indian regional and national YouTube news channels. \citet{feldman2021analyzing} fine-tuned GPT-2 on COVID-19 tweet corpora to mine user opinions. 




\begin{table}[!t]
\scriptsize
\centering
\begin{tabular}{ccc}
    \toprule
\textbf{Prompt}               & \textbf{Model}                        & \textbf{Top 5 Words}                                                                \\
    \midrule
\multirow{5}{*}{\makecell{Dr. Fauci \\is a}} & \textcolor{repubred}{Republican GPT-2}                      & \makecell{liar (2.96\%), joke (2.67\%), \\hero (2.13\%), doctor (1.62\%), \\great (1.61\%)}                     \\ \cmidrule{2-3} 
                              & \textcolor{demblue}{Democratic GPT-2}                      & \makecell{hero (10.36\%), true (3.63\%), \\national (2.08\%), physician (2.06\%), \\great (1.93\%)}   \\

\bottomrule
\end{tabular}
\caption{Top 5 words by odds for Republican and Democratic GPT-2 models, fine-tuned on partisan tweets. Dr. Fauci is suggested to be a ``hero'' by the GPT-2 model fine-tuned on Democratic tweets but a ``liar'' and ``joke'' by the GPT-2 model fine-tuned on Republican tweets.}
\label{tab:top_word_choice}
\end{table}

However, none of these studies fine-tune GPT-style language models on community data to probe community worldviews. In this work, we focus on Republican and Democratic Twitter communities and conduct a feasibility study using fine-tuned GPT-2 partisan language models to generate community responses and to predict community stance. As exemplified in Table \ref{tab:top_word_choice}, we observe clear partisan differences. In this sociopolitically fragmented society, our motivation is to provide a simple and flexible interface for people to probe each other's worldviews on topics of interest and to encourage constructive dialogue. We demonstrate through our experiments and analyses that the proposed method is a reliable tool to probe community opinions. The contribution of the work is as follows:

\begin{itemize}
    \item We present a simple \textsc{CommunityLM} framework based on GPT-2 language models to mine community insights by fine-tuning or training the model on community data. This study focuses on Democrat and Republican communities on Twitter but can be easily extended to probe insights from any community based on their public discourse or media diet\footnote{The source code of our paper is available at: \url{https://github.com/hjian42/CommunityLM}}.
    
    \item We use ANES questions as prompts and find that GPT-generated opinions are predictive of community stance towards public figures and groups. We experiment with 4 types of prompts and find that the fine-tuned \textsc{CommunityLM} with an ``X is the'' prompt outperforms all the baselines (including pre-trained GPT-3 Curie) in predicting community stance.
    
    
    
    
    \item We analyze the errors made by community language models and demonstrate the capability of the models to probe community preferences towards public figures by ranking.
    
    
\end{itemize}




\section{Partisan Twitter Data}


We construct a Twitter dataset containing 4.7M tweets (100M word tokens) by Republican and Democrat communities respectively. We first sampled 1M active U.S. Twitter users before and after the 2020 presidential election. We adapt the standard method \cite{volkova2014inferring,demszky2019analyzing} to estimate their political affiliation from the political accounts they follow and collect tweets of Republican and Democratic Twitter users from 2019-01-01 to 2020-04-10.  We pick this period because the ANES 2020 survey was collected between 2020-04-10 to 2020-04-18 and we want to ensure the Twitter training data does not leak information beyond 2020-04-10. We subsample 4.7M tweets from each side to achieve a balanced set and use \citet{nguyen-etal-2020-bertweet}'s tweet tokenizer for data processing. Details are described as follows.

\textbf{U.S. Twitter User Sampling.} We first sample a subset of active Twitter users from the ``decahose'', Twitter's 10\% sample of tweets. We define active U.S. users as those who posted at least 10 original tweets before and after the 2020 presidential election period (2020-07-01 to 2021-06-31). We then use  Litecoder\footnote{\url{https://github.com/social-machines/litecoder}} to extract user locations from their profile location strings and filter out users not based in the U.S. We construct the follow graph of the resulting set of 1,074,650 Twitter users.

\textbf{Partisan Assignment.} We follow previous studies \cite{volkova2014inferring,demszky2019analyzing} to estimate the party affiliation of Twitter users from the political accounts they follow. Specifically, we update the list of Twitter handles of US politicians from \citet{demszky2019analyzing} by adding current federal officeholders and governors from Ballotpedia\footnote{We update American politicians from \url{https://ballotpedia.org/List\_of\_current\_members\_of_the\_U.S.\_Congress} and \url{https://ballotpedia.org/Governor\_(state\_executive\_office)}}. The final list has 457 Republican and 473 Democratic politician Twitter handles. To identify committed partisan users, we adopt the following rules: a user is labeled as a Democrat if they followed no fewer than 6 Democratic politicians and no Republican politician from the list in February 2022, whereas a person is labeled as a Republican if they followed no fewer than 2 Republican politicians and no Democratic politicians. We choose these thresholds because there are 69\% Democratic users and 26\% Republican users on Twitter\footnote{\url{https://www.pewresearch.org/politics/2020/10/15/differences-in-how-democrats-and-republicans-behave-on-twitter}} (2.65:1). This step predicts 182,788 Democratic-leaning and 72,186 Republican-leaning users (2.53:1). 

\textbf{Tweet Pre-processing.} We use the tweet tokenizer from \citet{nguyen-etal-2020-bertweet} to process all the data. This tokenizer converts user mentions and
web/url links into special tokens @USER and HTTPURL. We delete HTTPURL from the tweets because it does not contain useful community information. We do not lower the case but filter out tweets with less than 10 tokens, producing 7,554,409 Democratic and 4,759,441 Republican tweets. We randomly sample from Democratic tweets to ensure both partisan communities have the same number of 4,759,441 tweets for training language models to ensure a fair community model comparison.



\section{Framework}

We present a simple \textsc{CommunityLM} framework which adapts GPT-style language models to mine community insights. This framework consists of four steps: (1) fine-tune or train GPT language models on community data, (2) design prompts based on survey questions, (3) generate community responses with language models, (4) aggregate community stance based on responses.


\subsection{Model Training and Fine-tuning}

We pick GPT-2 with 124M parameters and experiment with two training strategies on the partisan community data: (1) fine-tune a pre-trained GPT-2 model, (2) train a GPT-2 model from scratch. For both settings, we adopt training epoch 10 and batch size 24 on Nvidia GeForce GTX 1080 12GB. The greedy decoding is used for GPT-2. Otherwise, we use the default training parameters\footnote{\url{https://github.com/huggingface/transformers/blob/main/examples/pytorch/language-modeling/run_clm.py}}. The pre-trained GPT-2 model was released in February 2019, trained on data that cuts off at the end of 2017. We also use GPT-3 Curie as one of our baselines, which used training data up to Oct 2019. Therefore, neither pre-trained model used any data beyond the the start date of the ANES survey.

We adopt 10 epochs because GPT-2 was not pre-trained on the Twitter domain and had a steady loss decrease across all epochs. We checked all synthetic tweets (lowercased) generated by the fine-tuned GPT-2 with ``X is/are the''. The percentages of synthetic tweets appearing in training data are 64.93\% and 69.56\% for Republican and Democratic models. For researchers who want to adapt our approach with a lower repetition rate, we suggest moving away from the greedy decoding algorithm and reducing the epoch number.



\subsection{Prompt Design}

We design discrete prompts based on survey questions to probe community insights towards public figures and groups. The \textbf{American National Election Studies (ANES)} are academically-run national surveys of voters in the United States. We adopt the ANES 2020 Exploratory Testing Survey\footnote{\url{https://electionstudies.org/data-center/2020-exploratory-testing-survey/}} conducted between April 10, 2020 and April 18, 2020 on 3,080 adult citizens from across the United States, because this survey captures recent political changes in the US. We adapt all  30 questions from ``FEELING THERMOMETERS'' section of the ANES survey, which asks participants to rate people or groups from 0 (``not favorable'') to 100 (``favorable'') with the question ``How would you rate \_\_\_\_?'' The questions cover 30 items in two categories (a) \textbf{16 people}: Donald Trump, Barack Obama, Joe Biden, Elizabeth Warren, Bernie Sanders, Pete Buttigieg, Kamala Harris, Amy Klobuchar, Mike Pence, Andrew Yang, Nancy Pelosi, Marco Rubio, Alexandria Ocasio-Cortez, Nikki Haley, Clarence Thomas, Dr. Anthony Fauci, (b) \textbf{14 groups:} blacks, whites, Hispanics, Asians, illegal immigrants, feminists, the \#MeToo movement, transgender people, socialists, capitalists, big business, labor unions, the Republican Party, the Democratic Party. For each item ``X'', we experiment with four types of discrete prompts: (1) ``X'', (2) ``X is/are'', (3) ``X is/are a'', (4) ``X is/are the''. These names are copied from the survey verbatim except for ``whites'', ``blacks'', ``Hispanics'', and ``Asians'' because ``whites'' and ``blacks'' also refer to other named entities such as ``Blacks Clothing Company'' and ``Whites TV shows''. Instead, we translate the names of these four groups into ``White people'', ``Black people'', ``Hispanic people'', ``Asian people''. We also provide the count number of each item in Appendix \ref{sec:appendix-keyword-counts}.


\subsection{Community Response Generation}

For each community, we use the corresponding language model to generate 1000 responses given the prompts. We use Hugging Face's \texttt{TextGenerationPipeline} and apply the same decoding strategy by setting do\_sample to true, temperature to 1.0, and max\_length  to 50. If one response contains multiple sentences, we use the first line in the response and remove the remaining tokens, because a response with multiple sentences may have mixed sentiments, making it hard to identify the overall sentiment.



\subsection{Community Stance Aggregation}

After response generation, we save them locally and compute the community stance for each prompt by aggregating the sentiment of the synthetic responses. Specifically, we use the state-of-the-art Twitter sentiment classifier ``cardiffnlp/twitter-roberta-base-sentiment-latest'' \cite{barbieri-etal-2020-tweeteval,loureiro2022timelms} on the SemEval-2017 benchmark \cite{rosenthal-etal-2017-semeval} to classify each generated response into -1 (``Negative''), 0 (``Neutral''), and 1 (``Positive''). We take the average sentiment of the generated responses as the community's stance score towards the person or group. We also show the results of a popular lexicon-based sentiment classifier VADER in Appendix \ref{sec:appendix-classifier}. 

\section{Evaluation}

\textbf{Task Formulation.} The ANES survey has self-reported party affiliation from participants. We use responses from Republican and Democratic participants and calculate their average ratings towards each of 30 items (persons and groups). These average ratings are provided in Appendix \ref{sec:appendix-ratings}. If the average rating of Republican participants is higher than that of Democratic participants toward one item (e.g., Joe Biden), it is labeled as ``R''. Otherwise, the item is labeled as ``D''. 70\% items are labeled ``D'' and 30\% ``R''. The 9 items with ``R'' label are Donald Trump, Mike Pence, Marco Rubio, Nikki Haley, Clarence Thomas, whites, capitalists, big business, and the Republican Party. The task asks a model to predict which community is more favorable towards an item. To address the data imbalance, we prefer weighted F1 to accuracy as a measure of model performance.

\begin{table*}[ht!]
\small
    \centering
\begin{tabular}{lcccccc}
    \toprule
    \textbf{Model} & \textbf{Prompt} & \textbf{Accuracy}  & \textbf{Weighted F1} \\
    \midrule
    Frequency Model & --- &  53.33 &  54.50 \\
    Keyword Retrieval (Full) & --- &  86.67 & 87.00 \\
    Keyword Retrieval (Surname) & --- &  93.33 & 93.33 \\
    \midrule
    Pre-trained GPT-2 & ``[\textsc{Context}] + X'' &  74.00$\pm$2.79 & 66.52$\pm$5.56 \\
    Pre-trained GPT-2 & ``[\textsc{Context}] + X is/are'' &  72.00$\pm$1.83 & 64.63$\pm$2.35 \\
    Pre-trained GPT-2 & ``[\textsc{Context}] + X is/are a'' &  75.33$\pm$1.83 & 68.47$\pm$3.35 \\
    Pre-trained GPT-2 & ``[\textsc{Context}] + X is/are the'' &  77.33$\pm$2.79 &  74.71$\pm$3.22 \\
    \midrule
    Pre-trained GPT-3 Curie & ``[\textsc{Context}] +  X'' &  83.33 & 83.88 \\
    Pre-trained GPT-3 Curie & ``[\textsc{Context}] +  X is/are'' &  93.33 & 93.50 \\
    Pre-trained GPT-3 Curie & ``[\textsc{Context}] +  X is/are a'' &  83.33 & 83.88 \\
    Pre-trained GPT-3 Curie & ``[\textsc{Context}] +  X is/are the'' &  83.33 & 84.02 \\
    \midrule
    Trained \textsc{CommunityLM} & ``X'' &  90.00$\pm$0.00 & 89.63$\pm$0.27 \\
    Trained \textsc{CommunityLM} & ``X is/are'' &  90.00$\pm$0.00 & 89.82$\pm$0.00 \\
    Trained \textsc{CommunityLM} & ``X is/are a'' &  86.00$\pm$1.49 & 86.25$\pm$1.50 \\
    Trained \textsc{CommunityLM} & ``X is/are the'' &  90.67$\pm$2.79 & 90.49$\pm$2.68 \\
    \midrule
    Fine-tuned \textsc{CommunityLM} & ``X'' &  84.67$\pm$2.98 & 84.46$\pm$3.18 \\
    Fine-tuned \textsc{CommunityLM} & ``X is/are'' &  96.00$\pm$2.79 & 96.00$\pm$2.79 \\
    Fine-tuned \textsc{CommunityLM} & ``X is/are a'' &  91.33$\pm$1.83 & 90.83$\pm$2.05 \\
    Fine-tuned \textsc{CommunityLM} & ``X is/are the'' &  \textbf{97.33$\pm$1.49} &  \textbf{97.29$\pm$1.52} \\
    \bottomrule
    \end{tabular}
   \caption{Performance of different approaches in accuracy to predict which community is more favorable towards 30 persons or groups from the ANES survey. Approaches based on GPT-2 are repeated five times to compute the average and standard deviation. GPT-3 is only run once for cost concern. Frequency Model and Keyword Retrieval methods are deterministic. The weighted average F1 is used because of data imbalance.}
    \label{tab:metrics}
\end{table*}

\textbf{Baselines}. We evaluate the performance of trained and fine-tuned \textsc{CommunityLM} (GPT-2) against 4 baselines. The first baseline is Frequency Model which counts the frequency of an item's name in each community's data and classifies the community with higher word frequency to be the label. The second baseline is Keyword Retrieval which uses keywords to retrieve tweets containing the keywords from each community's data, computes the average community stance, and selects the community with a higher stance score. Keyword Retrieval (full) means using the full names as keywords and Keyword Retrieval (surname) means using the surname of people. The third and fourth baselines use pre-trained GPT-2 and pre-trained GPT-3 Curie respectively. ``[\textsc{CONTEXT}]'' is a preceding context ``As a Democrat/Republican, I think'', which is concatenated with the prompts to generate partisan responses on each item. We compute the average community stance on 1000 synthetic responses and pick the community with a higher average stance score. It is noted that we also fine-tune or train GPT-2 on the aggregate partisan tweets and show their results in Appendix \ref{sec:appendix-combined-data}.


\textbf{Overall Performance.} First, we observe that fine-tuned \textsc{CommunityLM} with ``X is the'' prompt achieves the best performance in both accuracy (97.33\%) and weighted F1-score (97.29\%) on the task. The same model's performance is sensitive to the prompt design and the longest prompt out of the four seems to work the best. ``X'' alone is bad, because it will result in many responses like ``X @USER'', ``X???'', ``X.'', which are common Twitter posts and are too short to interpret their attitudes. Second, fine-tuned \textsc{CommunityLM} outperforms trained \textsc{CommunityLM} from scratch. It indicates that pre-training GPT-2 is helpful, probably because pre-training injects the general knowledge about the named entities into GPT-2. Third, we find that Keyword Retrieval (surname) is a strong baseline in both accuracy (93.33\%) and F1 (93.33\%), but its performance is also sensitive to the selection of keywords. As we see, the weighted F1 performance of Keyword Retrieval (full), which uses a strict full name matching (e.g., ``Joe Biden''), drops to 87.00\%. In contrast, language models are able to learn the associations between different names for the same person and generalize without worrying about name forms. Last, fine-tuned \textsc{CommunityLM} outperforms pre-trained GPT-2 and GPT-3 baselines. It is worth noting that the performance of pre-trained GPT-3 Curie is consistently better than pre-trained GPT-2. GPT-3 with the ``X is/are'' prompt achieves the same score as the Keyword Retrieval (surname) baseline. 

\textbf{Error Analysis.} The rule-based Keyword Retrieval (surname) baseline misses ``illegal immigrants'' and ``big business''.  The fine-tuned \textsc{CommunityLM} with ``X is/are the'' misses ``White people''. The pre-trained GPT-3 with ``X is/are'' prompt misses ``Dr. Anthony Fauci'' and ``Asian people'. It is interesting because the top 5 items with the closest average rating gap between ANES partisan participants are Asian people (5.5\%), White people (5.9\%), Hispanic people (7.7\%), Dr. Anthony Fauci (8.4\%), and Black people (9.7\%). 

\begin{figure}[htb]
    \centering 
\begin{subfigure}{0.24\textwidth}
  \includegraphics[width=\linewidth]{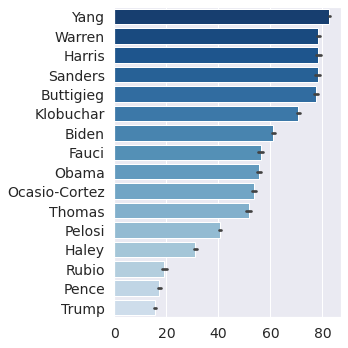}
  \caption{Democratic ranking}
  \label{fig:1}
\end{subfigure}\hfil 
\begin{subfigure}{0.24\textwidth}
  \includegraphics[width=\linewidth]{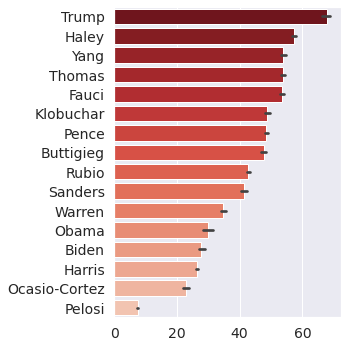}
  \caption{Republican ranking}
  \label{fig:2}
\end{subfigure}\hfil 
\caption{Left and right rankings of 16 public figures by their average stance scores calculated on synthetic tweets from their fine-tuned \textsc{CommunityLM} models.}
\label{fig:rank}
\end{figure}

\textbf{Ranking Public Figures.} We use the average community stance scores computed on the generated tweets from the fine-tuned \textsc{CommunityLM} model to rank 16 public figures for each community, hoping to understand how they perceive these people. In Figure \ref{fig:rank}, we observe that Republican politicians are rated poorly by the Democratic model and vice versa. Overall, the ratings from the Republican model are more negative than the Democratic model. Interestingly, we find that Andrew Yang is rated quite highly by both models, likely because of the sampling bias of Twitter. It is noted that ``Andrew Yang'' is also ranked 1st by the Democrat community and 3rd by the Republican community with the retrieval approach.



\section{Conclusion}

In this paper, we present a simple \textsc{CommunityLM} framework to evaluate the viability of fine-tuned GPT-2 community language models in mining community insights in the context of political polarization between Republicans and Democrats. We adopt ANES survey questions and experiment with four types of prompts to generate community responses through GPT-2, showing that generated opinions are predictive about which community is more favorable towards selected public figures and groups. Our results show that fine-tuned \textsc{CommunityLM} (GPT-2) outperforms the baseline methods. We analyze the model errors and run qualitative analyses to demonstrate that GPT-2 community language models can be used to rank public figures and probe word choices.



There are a few limitations in the current approach. First, language models can synthesize unreliable responses. Structured knowledge \cite{tacl_a_00360,yasunaga-etal-2021-qa} can be used to reduce nonsensical or unfaithful generation. Therefore, it is important that we use statistical patterns rather than individual synthesized tweets to draw conclusions \cite{feldman2021analyzing}. Second, language models are shown to be sensitive to prompt design in our experiments and are also vulnerable to negation and misprimed probes \cite{kassner-schutze-2020-negated}. In the future, we plan to develop a systematic approach to design effective prompts and evaluate the robustness of \textsc{CommunityLM}. Third, we focus on the classic red and blue polarization and do not consider a more fine-grained segmentation of U.S. politics. We hope to extend this work to study multiple sociopolitical communities in America and surface their unheard voices.

\section*{Ethical Considerations}
We propose a general framework to probe community insights and observe differences between the Democratic and Republican communities on Twitter. While we do not discuss how to react to these findings, the intention of our research is encourage people to escape from their echo chambers, hear voices from other communities, and engage in constructive communication.  One reasonable ethical concern is that by using a language model to predict community opinions, instead of asking individuals from the community directly, don't we risk erasing individual voices?  To that concern we would like to emphasize that our model is no substitute for deeper engagement with a community; as discussed in the limitation paragraph, the language model is just an entry point for understanding a community's perspective.  It serves to synthesize the points expressed by the speakers in the training data more effectively than we know how to do by hand.   Any automated or semi-automated prediction system risks misinterpreting or ``erasing'' an expressed opinion, and we show in our work that the simpler methods of doing so are more error-prone, and hence measurably more unfair than the proposed approach in the paper.

\section*{Acknowledgements}

We would like to thank anonymous reviewers for their helpful comments on our paper. We also want to thank Belén Carolina Saldías Fuentes, Will Brannon, Suyash Fulay, Wonjune Kang, Hope Schroeder, and Shayne Longpre for their discussion and feedback in the early stages of the project.

\bibliography{anthology,custom}
\bibliographystyle{acl_natbib}

\clearpage
\appendix
\section*{Appendix}

\section{Keyword counts}
\label{sec:appendix-keyword-counts}

The Keyword Retrieval baseline method retrieves tweets containing the keywords. 
Here we show the list of full and surname keywords and their counts in tables \ref{tab:full_counts} and \ref{tab:counts_surname}, respectively, for the Republican and Democratic tweets. 
For corresponding items between these two tables (e.g. ``Asian people'' in Table \ref{tab:full_counts} to ``Asian'' in Table \ref{tab:counts_surname}) there is a consistent increase in counts, especially for ``Asian people'' ``Anthony Fauci'', ``Hispanic people'', ``labor unions'', ``Clarence Thomas''. Some items in Table \ref{tab:counts_surname} might have too many counts. For example, we observe that ``Trump'' has 150,000+ counts in both partisan tweets, which can take a relatively long time for sentiment classifiers to run.

\begin{table}[h!]
\scriptsize
\centering
\begin{tabular}{|c|c|c|c|}
\hline
\textbf{Keyword}         & \textbf{Question} & \textbf{Dem} & \textbf{Repub} \\ \hline
Asian people             & ftasian           & 81           & 21             \\ \hline
Joe Biden                & ftbiden1          & 4177         & 5377           \\ \hline
big business             & ftbigbusiness     & 321          & 291            \\ \hline
Black people             & ftblack           & 3199         & 1278           \\ \hline
Pete Buttigieg           & ftbuttigieg1      & 982          & 521            \\ \hline
capitalists              & ftcapitalists     & 279          & 197            \\ \hline
the Democratic Party     & ftdemocraticparty & 2094         & 2646           \\ \hline
Anthony Fauci            & ftfauci1          & 102          & 85             \\ \hline
feminists                & ftfeminists       & 351          & 628            \\ \hline
Nikki Haley              & fthaley1          & 169          & 274            \\ \hline
Kamala Harris            & ftharris1         & 1711         & 1450           \\ \hline
Hispanic people          & fthisp            & 28           & 21             \\ \hline
illegal immigrants       & ftillegal         & 251          & 2233           \\ \hline
Amy Klobuchar            & ftklobuchar1      & 451          & 193            \\ \hline
labor unions             & ftlaborunions     & 68           & 27             \\ \hline
the \#MeToo movement     & ftmetoo           & 103          & 84             \\ \hline
Barack Obama             & ftobama1          & 684          & 929            \\ \hline
Alexandria Ocasio-Cortez & ftocasioc1        & 410          & 534            \\ \hline
Nancy Pelosi             & ftpelosi1         & 1467         & 3549           \\ \hline
Mike Pence               & ftpence1          & 911          & 502            \\ \hline
the Republican Party     & ftrepublicanparty & 1681         & 838            \\ \hline
Marco Rubio              & ftrubio1          & 166          & 132            \\ \hline
Bernie Sanders           & ftsanders1        & 4572         & 2711           \\ \hline
socialists               & ftsocialists      & 627          & 2697           \\ \hline
Clarence Thomas          & ftthomas1         & 157          & 132            \\ \hline
transgender people       & fttransppl        & 165          & 38             \\ \hline
Donald Trump             & fttrump1          & 8501         & 5479           \\ \hline
Elizabeth Warren         & ftwarren1         & 3132         & 1897           \\ \hline
White people             & ftwhite           & 3625         & 1862           \\ \hline
Andrew Yang              & ftyang1           & 585          & 249            \\ \hline
\end{tabular}
\caption{Counts of full names for each person and group in Republican and Democratic tweets.}
\label{tab:full_counts}
\end{table}

\begin{table}[h!]
\scriptsize
\centering
\begin{tabular}{|c|c|c|c|}
\hline
\textbf{Keyword}  & \textbf{Question} & \textbf{Dem} & \textbf{Repub} \\ \hline
Asian             & ftasian           & 2961         & 1917           \\ \hline
Biden             & ftbiden1          & 26558        & 21748          \\ \hline
big business      & ftbigbusiness     & 321          & 291            \\ \hline
Black people      & ftblack           & 3199         & 1278           \\ \hline
Buttigieg         & ftbuttigieg1      & 3514         & 1348           \\ \hline
capitalist        & ftcapitalists     & 1393         & 941            \\ \hline
Democratic Party  & ftdemocraticparty & 2677         & 3611           \\ \hline
Fauci             & ftfauci1          & 931          & 1219           \\ \hline
feminist          & ftfeminists       & 1686         & 1470           \\ \hline
Haley             & fthaley1          & 531          & 712            \\ \hline
Harris            & ftharris1         & 6753         & 5416           \\ \hline
Hispanic          & fthisp            & 1173         & 1693           \\ \hline
illegal immigrant & ftillegal         & 312          & 2815           \\ \hline
Klobuchar         & ftklobuchar1      & 1958         & 584            \\ \hline
labor union       & ftlaborunions     & 110          & 47             \\ \hline
\#MeToo movement  & ftmetoo           & 114          & 102            \\ \hline
Obama             & ftobama1          & 15390        & 33105          \\ \hline
Ocasio-Cortez     & ftocasioc1        & 751          & 1792           \\ \hline
Pelosi            & ftpelosi1         & 5985         & 15844          \\ \hline
Pence             & ftpence1          & 5818         & 3021           \\ \hline
Republican Party  & ftrepublicanparty & 2251         & 1079           \\ \hline
Rubio             & ftrubio1          & 508          & 502            \\ \hline
Sanders           & ftsanders1        & 16001        & 6568           \\ \hline
socialist         & ftsocialists      & 3182         & 12606          \\ \hline
Thomas            & ftthomas1         & 2316         & 3348           \\ \hline
transgender       & fttransppl        & 1309         & 1469           \\ \hline
Trump             & fttrump1          & 188170       & 150589         \\ \hline
Warren            & ftwarren1         & 18954        & 6969           \\ \hline
White people      & ftwhite           & 3625         & 1862           \\ \hline
Yang              & ftyang1           & 4443         & 1433           \\ \hline
\end{tabular}
\caption{Counts of surname names for each person and group in Republican and Democratic tweets.}
\label{tab:counts_surname}
\end{table}

\section{What are the average ratings between partisan participants in ANES survey?}
\label{sec:appendix-ratings}

We also compute and show in Table \ref{tab:rating-anes} the average ratings from Republican and Democratic participants towards each person or group. For most items, we observe quite large rating gaps between the partisans. But the top 5 items with the closest average rating gap between partisans are ``Asian people'' (5.5\%), ``White people'' (5.9\%), ``Hispanic people'' (7.7\%), ``Dr. Anthony Fauci'' (8.4\%), ``Black people'' (9.7\%). These items have very close ratings and we confirm in our error analysis that they are also challenging to the GPT-2 models. It is worth noting that the survey was done in early 2020 and at that time ``Dr. Fauci'' as a topic was not as divisive as it is today on Twitter.

\begin{table}[h!]
\scriptsize
\centering
\begin{tabular}{|c|c|c|c|}
\hline
\textbf{Question} & \textbf{Item}            & \textbf{Dem} & \textbf{Repub} \\ \hline
ftasian           & Asian people             & 68.95             & 63.44               \\ \hline
ftwhite           & White people             & 71.25             & 77.16               \\ \hline
fthisp            & Hispanic people          & 71.27             & 63.60               \\ \hline
ftfauci1          & Dr. Anthony Fauci        & 66.67             & 58.28               \\ \hline
ftblack           & Black people             & 76.22             & 66.51               \\ \hline
ftrubio1          & Marco Rubio              & 31.52             & 43.01               \\ \hline
ftcapitalists     & capitalists              & 46.68             & 60.53               \\ \hline
ftbigbusiness     & big business             & 43.14             & 57.85               \\ \hline
ftlaborunions     & labor unions             & 60.67             & 44.87               \\ \hline
fthaley1          & Nikki Haley              & 29.86             & 47.07               \\ \hline
ftthomas1         & Clarence Thomas          & 29.95             & 48.63               \\ \hline
ftyang1           & Andrew Yang              & 49.28             & 29.19               \\ \hline
ftklobuchar1      & Amy Klobuchar            & 50.04             & 22.17               \\ \hline
ftfeminists       & feminists                & 61.97             & 33.92               \\ \hline
fttransppl        & transgender people       & 63.22             & 35.06               \\ \hline
ftsocialists      & socialists               & 54.00             & 24.11               \\ \hline
ftillegal         & illegal immigrants       & 56.17             & 26.25               \\ \hline
ftmetoo           & the \#MeToo movement     & 63.74             & 32.73               \\ \hline
ftbuttigieg1      & Pete Buttigieg           & 52.79             & 21.66               \\ \hline
ftharris1         & Kamala Harris            & 52.12             & 18.63               \\ \hline
ftocasioc1        & Alexandria Ocasio-Cortez & 50.60             & 16.49               \\ \hline
ftwarren1         & Elizabeth Warren         & 59.84             & 20.46               \\ \hline
ftbiden1          & Joe Biden                & 66.50             & 24.40               \\ \hline
ftsanders1        & Bernie Sanders           & 63.77             & 20.50               \\ \hline
ftpelosi1         & Nancy Pelosi             & 61.76             & 16.10               \\ \hline
ftdemocraticparty & the Democratic Party     & 71.24             & 24.34               \\ \hline
ftpence1          & Mike Pence               & 24.09             & 71.12               \\ \hline
ftrepublicanparty & the Republican Party     & 25.02             & 74.47               \\ \hline
ftobama1          & Barack Obama             & 81.29             & 29.99               \\ \hline
fttrump1          & Donald Trump             & 17.66             & 77.83               \\ \hline
\end{tabular}
\caption{Average rating of each item (person or group) from Republican and Democratic participants in the ANES survey.}
\label{tab:rating-anes}
\end{table}

\section{How well does the system perform using a lexicon-based sentiment classifier?}
\label{sec:appendix-classifier}

In the main paper, we use a state-of-the-art pre-trained BERT Twitter sentiment classifier to classify tweets. Some researchers may be concerned that neural sentiment models may learn and reflect biases in the training data and prefer using lexicon-based approaches. Therefore, we also use VADER \cite{hutto2014vader}\footnote{\url{https://github.com/cjhutto/vaderSentiment}}, a popular rule-based model for sentiment analysis of social media texts, and report the performance of our models with VADER in Table \ref{tab:vader}. Overall, we show that these models perform slightly worse with VADER, but we still see that fine-tuned \textsc{CommunityLM} with ``X is the'' perform the best (93.33\%) out of these models. This performance is on par with the Keyword Retrieval (surname) approach. We conjecture that using prompts like ``X is the'' creates many synthetic tweets with only sentiment-neutral lexical items (e.g., ``big business is the future'') which the lexicon-based VADER is not able to classify as ``positive''. The BERT sentiment classifier, however, performs better at representing the overall semantics of the sentence and therefore is preferred in our framework.

\begin{table}[ht!]
\tiny
    \centering
\begin{tabular}{lccc}
    \toprule
    \textbf{Model} & \textbf{Prompt} & \textbf{Accuracy} \\
    \midrule
    Keyword Retrieval (Full) & --- &  76.67 \\
    Keyword Retrieval (Surname) & --- &  93.33 \\
    \midrule
    Pre-trained GPT-2 & ``[\textsc{Context}] + X'' &  76.67$\pm$0.00 \\
    Pre-trained GPT-2 & ``[\textsc{Context}] + X is/are'' &  76.00$\pm$1.49 \\
    Pre-trained GPT-2 & ``[\textsc{Context}] + X is/are a'' &  78.67$\pm$1.83 \\
    Pre-trained GPT-2 & ``[\textsc{Context}] + X is/are the'' &  74.67$\pm$5.06 \\
    \midrule
    Trained \textsc{CommunityLM} & ``X'' &  91.33$\pm$3.80 \\
    Trained \textsc{CommunityLM} & ``X is/are'' &  84.67$\pm$3.80 \\
    Trained \textsc{CommunityLM} & ``X is/are a'' &  82.00$\pm$3.80 \\
    Trained \textsc{CommunityLM} & ``X is/are the'' &  93.33$\pm$4.08 \\
    \midrule
    Fine-tuned \textsc{CommunityLM} & ``X'' &  92.00$\pm$1.83 \\
    Fine-tuned \textsc{CommunityLM} & ``X is/are'' &  92.67$\pm$1.49 \\
    Fine-tuned \textsc{CommunityLM} & ``X is/are a'' &  91.33$\pm$1.83 \\
    Fine-tuned \textsc{CommunityLM} & ``X is/are the'' &  93.33$\pm$2.36 \\
    \bottomrule
    \end{tabular}
   \caption{Performance of different approaches with VADER in predicting which community is more favorable towards 30 persons or groups from the ANES survey. Approaches based on GPT-2 are repeated five times to compute the average and standard deviation.}
    \label{tab:vader}
\end{table}

\section{Is fine-tuning or training GPT-2 on combined Twitter data performing better than pre-trained GPT-2?}
\label{sec:appendix-combined-data}

In the main paper, we use pre-trained GPT-2 and GPT-3 to predict the community stance. In addition, we also experimented with training and fine-tuning GPT-2 on the combined Twitter corpus (Republican and Democratic tweets). By contrast with \textsc{CommunityLM}, which fine-tunes two GPT-2 models on partisan Twitter data, in this variant we only train or fine-tune one GPT-2 model on the aggregate of the partisan tweets. Similar to what we did in the pre-trained GPT-2 setting, we use [CONTEXT]+prompt to generate responses. The results are quite interesting, because the performance of the resulting models are worse than the pre-trained GPT-2, even below the majority baseline of 70\%. We conjecture that this is because the combined data of partisan tweets neutralizes the sentiment that the models were supposed to learn towards the public figures and groups. 

\begin{table}[ht!]
\tiny
    \centering
\begin{tabular}{lccc}
    \toprule
    \textbf{Model} & \textbf{Prompt} & \textbf{Accuracy} \\
    \midrule
    Trained GPT-2 (combined) & ``[\textsc{Context}] + X'' &  48.67$\pm$8.37 \\
    Trained GPT-2 (combined) & ``[\textsc{Context}] + X is/are'' &  50.67$\pm$7.23 \\
    Trained GPT-2 (combined) & ``[\textsc{Context}] + X is a'' &  47.33$\pm$2.79 \\
    Trained GPT-2 (combined) & ``[\textsc{Context}] + X is the'' &  55.33$\pm$8.37 \\
    \midrule
    Fine-tuned GPT-2 (combined) & ``[\textsc{Context}] + X'' &  53.33$\pm$4.08 \\
    Fine-tuned GPT-2 (combined) & ``[\textsc{Context}] + X is/are'' &  50.67$\pm$3.65 \\
    Fine-tuned GPT-2 (combined) & ``[\textsc{Context}] + X is a'' &  52.67$\pm$2.79 \\
    Fine-tuned GPT-2 (combined) & ``[\textsc{Context}] + X is the'' &  38.00$\pm$8.37 \\
    \bottomrule
    \end{tabular}
   \caption{Performance of trained and fine-tuned GPT-2 on combined Twitter data in accuracy to predict which community is more favorable towards 30 persons or groups from the ANES survey. Experiments are repeated five times to compute the average and standard deviation.}
    \label{tab:combine_exp}
\end{table}

\end{document}